\documentclass[a4paper,11pt]{article}
\usepackage{pos}

\title{How the nonequilibrium initial stages influence medium-induced radiation}
\author*[a]{Florian Lindenbauer}

\affiliation[a]{MIT Center for Theoretical Physics – a Leinweber Institute,
Massachusetts Institute of Technology, \\  77 Massachusetts Avenue, Cambridge, MA 02139, USA}

\emailAdd{flindenb@mit.edu}

\abstract{We compute the energy loss and gluon radiation spectrum of an energetic parton traversing a nonequilibrium QCD plasma. Using QCD kinetic theory simulations of a Bjorken expanding plasma, we extract the small-distance form of the jet-medium potential, which we use as input to calculate the radiation spectrum within the improved opacity expansion. We compare our results for a time-evolving nonequilibrium plasma with those of a Landau-matched thermal plasma and identify differences even for large systems.}

\notes{\note{    This work is funded in part by the Austrian Science Fund (FWF) under Grant DOI 10.55776/P34455, and Grant DOI 10.55776/J4902. FL was a recipient of a DOC Fellowship of the Austrian Academy of Sciences at TU Wien (project 27203). We thank ECT* for support at the Workshop ``New jet quenching tools to explore equilibrium and non-equilibrium dynamics in heavy-ion collisions'' during which this work was initiated. The results in this paper have been achieved in part using the Austrian Scientific Computing (ASC) infrastructure, project 71444. For the purpose of open access, the author has applied a CC BY public copyright license to any Author Accepted Manuscript (AAM) version arising from this submission.
}}

\FullConference{13th International Conference on Hard and Electromagnetic Probes of High-Energy Nuclear Collisions (HardProbes2026)\\
21-26 June 2026\\
Nashville, Tennessee, USA\\}

\newcommand{\pdv}[2][]{\frac{\partial{#1}}{\partial{#2}}}
\newcommand{\vb}[1]{\mathbf{#1}}
\newcommand{\dv}[2][]{\frac{\dd{#1}}{\dd{#2}}}
\newcommand{\dd}[2][]{\mathrm d^{#1}{#2}\,}

\begin{document}
\hfill MIT-CTP/6172
\maketitle

\section{Introduction: Kinetic theory and jet quenching}
\begin{figure}
    \centering
    \centerline{    \includegraphics[width=0.33\linewidth]{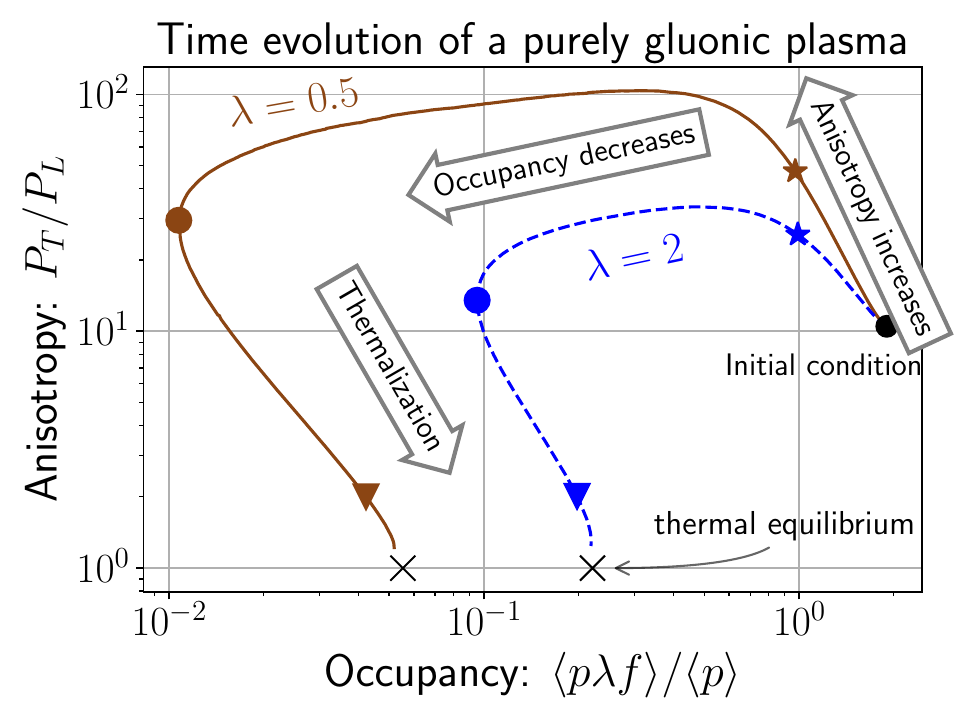}
\includegraphics[width=0.33\linewidth]{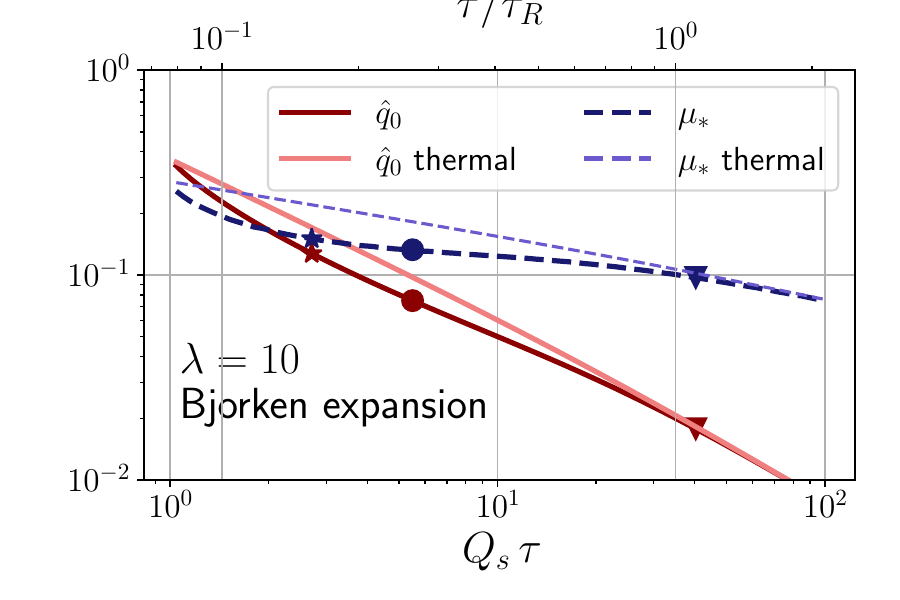}
\includegraphics[width=0.33\linewidth]{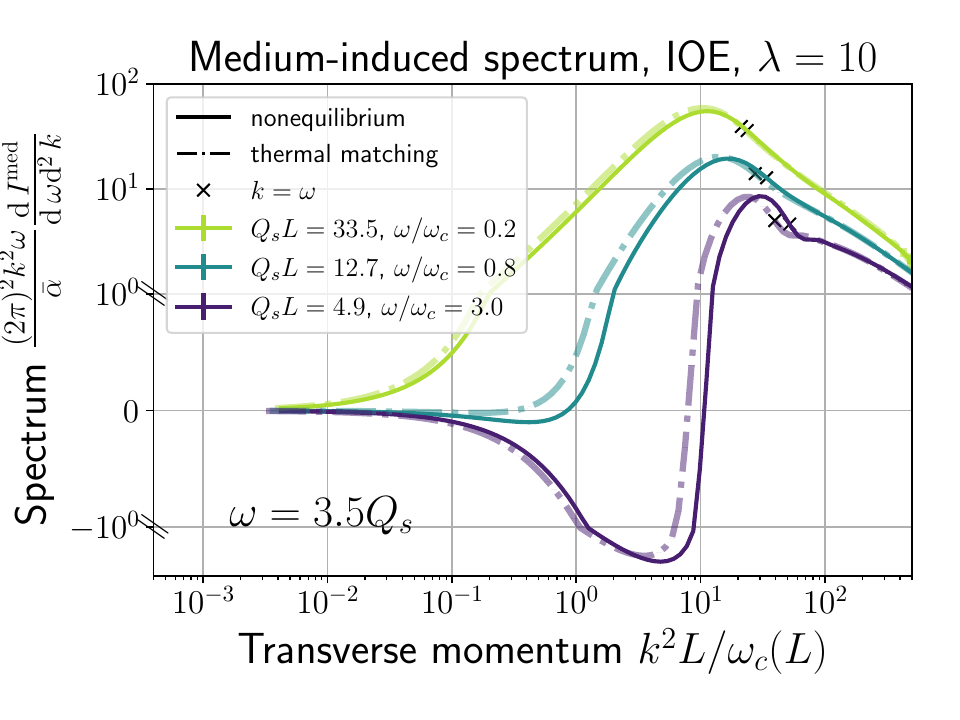}
    }
    \caption{(Left): Time-evolution of an initially overoccupied Yang-Mills plasma undergoing boost-invariant Bjorken expansion, shown in the anisotropy-occupancy plane. (Center): Parameters $\hat q_0(\tau)$ and $\mu_*(\tau)$ during the bottom-up evolution compared to instantaneous thermal values. (Right): Gluon emission spectrum calculated from the time-dependent parameters $\hat q_0(\tau)$ and $\mu_*(\tau)$. Figures from \cite{Lindenbauer:2025ael, Barata:2025agq}.}
    \label{fig:kinetictheory}
\end{figure}

The QCD plasma of deconfined quarks and gluons created in the aftermath of a heavy-ion collision is initially very far from equilibrium. While hydrodynamics can be successfully used to model the late-time evolution of this plasma, the early far-from-equilibrium stages are not captured by a fluid description. 
In this work, we employ QCD effective kinetic theory \cite{Arnold:2002zm}, which is valid in the weak-coupling limit of QCD, and commonly applied at larger, more realistic couplings as a model in lieu of a full first-principles method.
Within this method, the thermalization of a QCD plasma and the approach to hydrodynamics have been studied extensively \cite{Kurkela:2014tea, Kurkela:2015qoa, Fu:2021jhl, Kurkela:2018xxd, Du:2020zqg, Kurkela:2018wud, Boguslavski:2024kbd, BarreraCabodevila:2025ogv}. A weakly-coupled initially overoccupied gluon plasma follows the bottom-up evolution \cite{Baier:2000sb}, consisting of an initially overoccupied and later underoccupied stage. Numerical results for these simulations are depicted in the left panel of Fig.~\ref{fig:kinetictheory}, obtained from a numerical solution of the Boltzmann equation \cite{Arnold:2002zm, Mueller:1999pi, kurkela_2023_10409474}
\begin{align}
	\pdv[f(\vec p,\tau)]{\tau}- \frac{p_z}{\tau} \pdv[f(\vec p,\tau)]{p_z} =-C[f(\vec p,\tau)] \, ,\label{eq:boltzmann_equation}
\end{align}
with an initial condition motivated by the earlier overoccupied Glasma state \cite{Lappi:2011ju}, as parameterized in \cite{Kurkela:2015qoa}. In addition, we also consider the time evolution of isotropic and spatially homogeneous toy models, which are initially over- or underoccupied. For these, the expansion term evolving the $p_z$ derivative in Eq.~\eqref{eq:boltzmann_equation} is missing.

To access these initial nonequilibrium stages experimentally, it has been suggested to measure the suppression and substructure of jets, particle showers originating from a highly energetic parton created in the initial hard collision process. This process is known as jet quenching, and many theoretical formalisms have been developed to describe and calculate the energy loss of jets \cite{Apolinario:2022vzg}. How jets evolve during the early stages has recently received increased attention \cite{Ipp:2020nfu, Carrington:2021dvw, Avramescu:2023qvv, Barata:2024xwy, Boguslavski:2024ezg, Altenburger:2025iqa, Zhou:2024ysb, Barata:2025zku}. In these proceedings, we calculate the gluon emission spectrum of an energetic parton using the \emph{Improved Opacity Expansion} (IOE) formalism \cite{Mehtar-Tani:2019ygg, Barata:2021wuf}, 
going beyond the often-used harmonic approximation by an expansion in logs. 

\section{Parameterizing the scattering potential}

The input for calculating this spectrum is a microscopic scattering potential $v(\vb x, t)$, 
\begin{align}
	v(\vb x,t) = \int \frac{\mathrm{d}^2{\vb q}}{(2\pi)^2}(1-e^{i\vb x\cdot \vb q}) C(\vb q)\simeq\frac{\hat q_0(t)}{4} \vb x^2 \log \frac{1}{\vb x^2 \mu_\ast^2(t)} +\mathcal{O}(\hat q_0 \vb x^4 \mu_\ast^2 ) \, , 
	\label{eq:potential_expansion}
\end{align}
which enters an effective two-dimensional Schrödinger equation that needs to be solved for calculating the spectrum \cite{Barata:2021wuf}.
This scattering potential $v(\vb x,t)$ entering Eq.~\eqref{eq:potential_expansion} has recently been obtained from QCD kinetic theory simulations \cite{Altenburger:2025iqa}. In Ref.~\cite{Altenburger:2025iqa, Barata:2025agq}, it was realized that the parameters $\hat q_0$ and $\mu_\ast$ in Eq.~\eqref{eq:potential_expansion}, which characterize the small-distance form of the potential, can be obtained from the logarithmic dependence of the jet quenching parameter 
\begin{align}
	\hat q( \Lambda_\perp, \tau) =\int_{|\vb q|<\Lambda_\perp} \frac{\mathrm{d}^2 \vb q}{(2\pi)^2}\vb q^2C(\vb q,\tau)\simeq a(\tau)\log\Lambda_\perp + b(\tau)
\end{align} on its transverse momentum cutoff $\Lambda_\perp$,
\begin{align}
	\hat q_0(\tau)=a(\tau), && \mu_\ast^2(\tau)=\frac{\exp\left(-\frac{b(\tau)}{\hat q_0(\tau)} + 2\gamma_E-2\right)}{4}.
\end{align}
The central panel of  Fig.~\ref{fig:kinetictheory} shows the extraction of the parameters $\hat q_0$ and $\mu_\ast$ as a function of time, compared to an instantaneous thermal system, defined as the thermal system with the same energy density (Landau matching). There, the parameters are\footnote{For larger couplings $\lambda \gtrsim 2$, the results of \cite{Arnold:2008vd} exhibit a small offset, and the numerical values from \cite{Boguslavski:2024jwr} are used, as explained in more detail in \cite{Barata:2025agq}.}
\begin{align}
    \hat q_0=T^3 \frac{N_CC_Rg^4\zeta(3)}{\pi^3}, && \mu_\ast^2&=\frac{m_D^2}{4}\exp\left[2\gamma_E-2-\left(\frac{\zeta(2)}{\zeta(3)}-1\right)(1-2\gamma_E)+\frac{\sigma_+}{\pi\zeta(3)}\right]\left(\frac{m_D}{2T}\right)^{2\zeta(2)/\zeta(3)-2},
\end{align}
where $\zeta$ is the Riemann Zeta function, $\gamma_E\approx 0.57722(\dots)$ is the Euler-Mascheroni constant, and $\sigma_+=0.386043817(\dots)$. 

Focusing on $\hat q_0$ (red curves), we observe that the Bjorken expanding plasma starts out with a $\hat q_0$ comparable to the thermal value, which quickly changes as $\hat q_0$ drops below the thermal reference value until approaching the equilibrium value from below. This is in nice agreement with the physical picture of bottom-up thermalization. There, the plasma is initially overoccupied, implying a larger $\hat q_0$ than in equilibrium, followed by an underoccupied stage, where $\hat q_0$ is smaller than in equilibrium. The fact that we can observe this underoccupied stage clearly in the behavior of $\hat q_0$ in the central panel of Fig.~\ref{fig:kinetictheory} will be a crucial ingredient for our interpretation of the spectrum.

\section{Obtaining the spectrum}
To obtain numerical results for the spectrum, we first solve a differential equation for obtaining the spectrum in the harmonic approximation \cite{Arnold:2008iy}, followed by an expansion around the harmonic solution \cite{Barata:2021wuf}. Anisotropic systems require a more complicated setup even in the harmonic case \cite{Barata:2024bqp, Boguslavski:2026wma}. Here, we approximate the system as isotropic in the plane transverse to the jet, which is also consistent with the current kinetic theory implementations \cite{kurkela_2023_10409474} (see \cite{, Lindenbauer:2025ctw} for how to go beyond isotropy in the transverse plane).
The analytic expressions for the spectrum are lengthy and can be found in \cite{Barata:2025agq}, and the implementation and data are publicly accessible \cite{lindenbauer_2025_17552870}.
\section{Results for the spectrum}
\begin{figure}
	\centerline{
    		\includegraphics[width=0.33\linewidth]{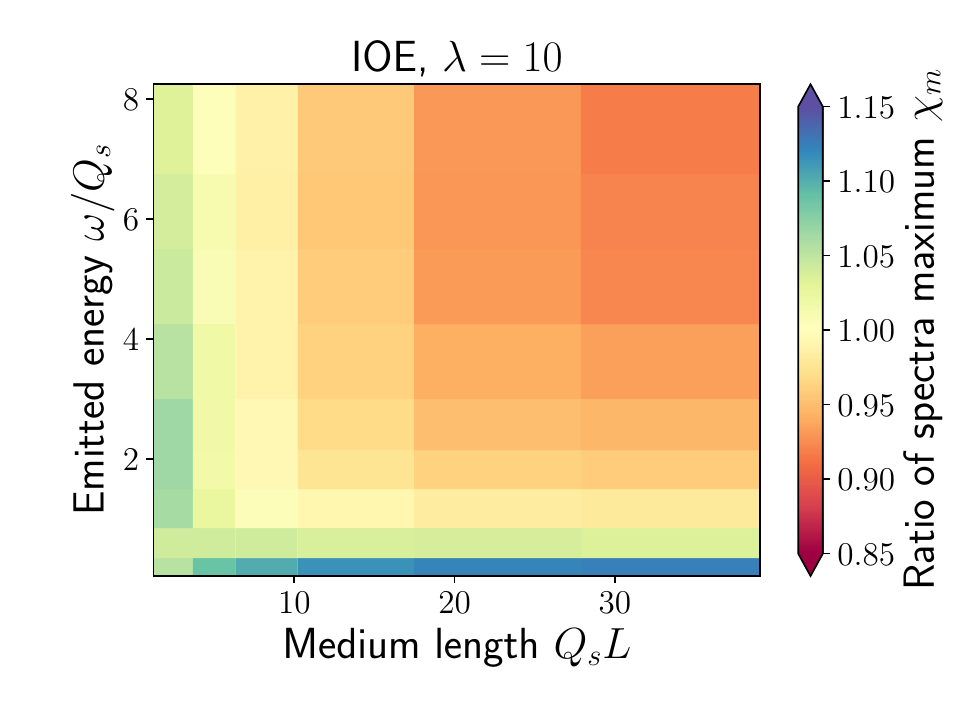}
            \includegraphics[width=0.33\linewidth]{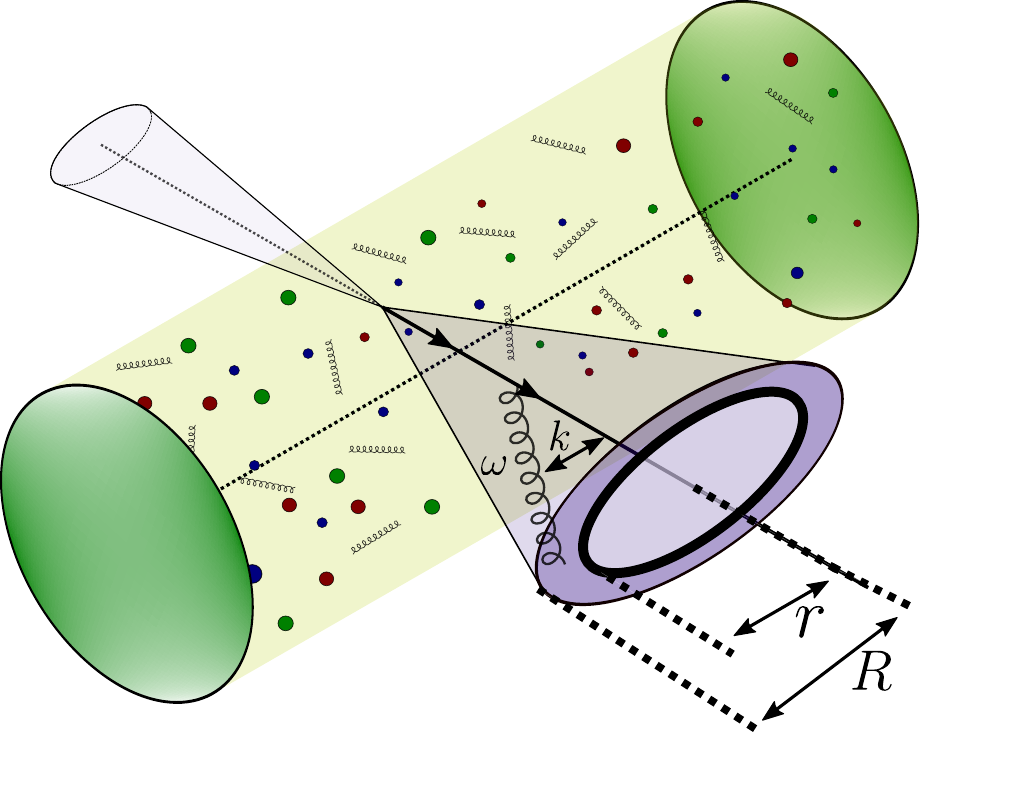}

		\includegraphics[width=0.33\linewidth]{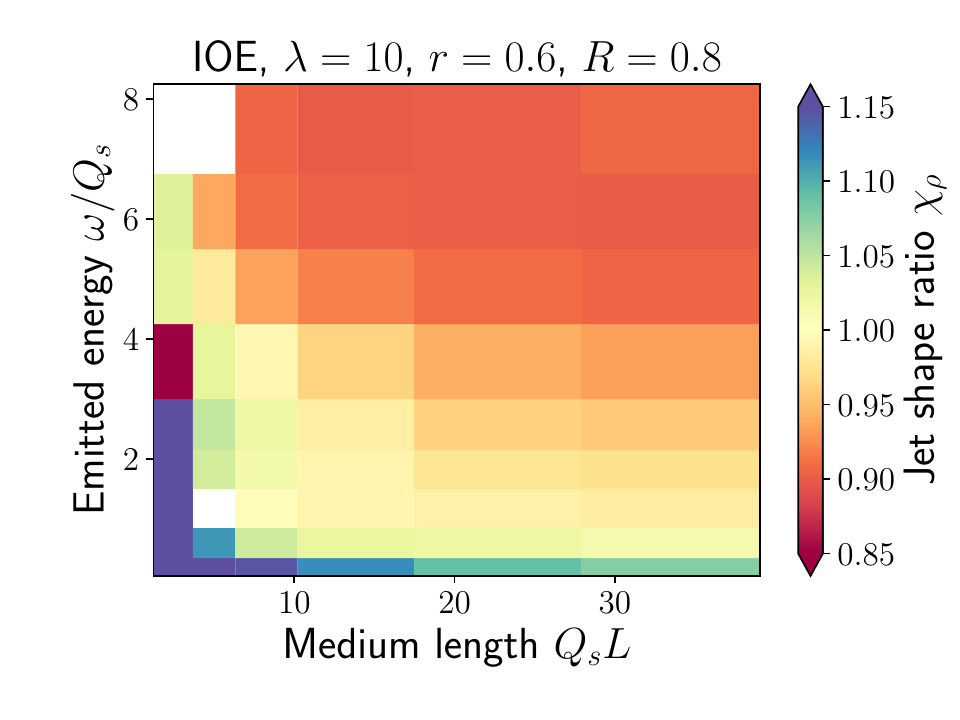}
	}
	\caption{(Left): Ratio of spectra maxima $\chi_m$ for different gluon energies $\omega$ and lengths $L$ for a Bjorken expanding plasma. (Center): Sketch of the jet shape observable. (Right): Ratio of jet shapes $\chi_\rho$ for different gluon energies $\omega$ and lengths $L$. Figures from \cite{Barata:2025agq}.}
	\label{fig:spectra}
\end{figure}
We show the resulting spectrum $\frac{\mathrm{d}I}{\mathrm{d\omega \mathrm{d^2k}}}$ in the right panel of Fig.~\ref{fig:kinetictheory}, comparing the spectrum from the nonequilibrium evolution (solid lines) to the thermal estimates (dash-dotted lines). To gain intuition for the differences in the spectra, we study systematic variations in the parameters, such as the gluon emission energy $\omega$ and the medium length $L$ (different colors), as shown systematically in Fig.~\ref{fig:spectra} (left panel). For a fixed emitted gluon energy $\omega$, we find that increasing the length leads to a reduction in the ratio $\chi_m(\omega, L)=\frac{\max_k\frac{k^2\mathrm{d}{I^\mathrm{noneq}}}{\mathrm{d}\omega \mathrm{d}^2{\vb k}}}{\max_k\frac{k^2\mathrm{d}{I^\mathrm{eq}}}{\mathrm{d}\omega \mathrm{d}^2{\vb k}}}
$, dropping from a value initially above unity to a value below unity. This can be understood, as anticipated before, as a result of the initially over- and then underoccupied stages during the bottom-up evolution. For isotropic toy models, we find that initially overoccupied systems are characterized by $\chi_m>1$, whereas the underoccupied plasma has $\chi_m<1$, which explains the behavior observed in the left panel of Fig.~\ref{fig:spectra}. 

A more physical way of characterizing the difference is to consider the jet shape
\begin{align}
    \dv[\rho(r,R)]{\omega}=\int_{r\omega}^{R\omega}\dd[2]{\vb {k}} \,\omega \frac{\dd I}{\dd\omega\dd[2]{\vb k}} \, ,
\end{align}
i.e., the energy deposited between radii $r$ and $R$, as sketched in the central panel of Fig.~\ref{fig:spectra}. The right panel of this figure depicts the ratio of these quantities $\chi_\rho(\omega,L; r,R)=\frac{\dv[\rho^\mathrm{noneq}(r,R)]{\omega}}{\dv[\rho^{\mathrm{eq}}(r,R)]{\omega}}$. It exhibits a similar behavior to the ratio of the maxima $\chi_m$, and demonstrates that particularly the underoccupied stage leaves a sizable imprint on this quantity.

\section{Conclusions}
Our results demonstrate that the nonequilibrium medium evolution in heavy-ion collisions influences the gluon emission spectrum of jets, and is not equivalent to replacing the nonequilibrium medium evolution with a thermal system with varying temperature. Future work is needed to study how these effects propagate through a more realistic phenomenological setup, and to compute more differential observables, which may be sensitive to the large initial plasma anisotropy.

\bibliographystyle{JHEP}
\bibliography{bib}

@article{Arnold:2008vd,
    author = "Arnold, Peter Brockway and Xiao, Wei",
    title = "{High-energy jet quenching in weakly-coupled quark-gluon plasmas}",
    eprint = "0810.1026",
    archivePrefix = "arXiv",
    primaryClass = "hep-ph",
    doi = "10.1103/PhysRevD.78.125008",
    journal = "Phys. Rev. D",
    volume = "78",
    pages = "125008",
    year = "2008"
}

@article{Barata:2025agq,
    author = "Barata, Jo{\~a}o and Boguslavski, Kirill and Lindenbauer, Florian and Sadofyev, Andrey V.",
    title = "{Jet quenching in out-of-equilibrium QCD matter}",
    eprint = "2511.07519",
    archivePrefix = "arXiv",
    primaryClass = "hep-ph",
    reportNumber = "CERN-TH-2025-140, MIT-CTP/5953",
    doi = "10.1103/q8bb-qlm8",
    journal = "Phys. Rev. D",
    volume = "113",
    number = "11",
    pages = "114009",
    year = "2026"
}

@article{Boguslavski:2024jwr,
    author = "Boguslavski, Kirill and Kurkela, Aleksi and Lappi, Tuomas and Lindenbauer, Florian and Peuron, Jarkko",
    title = "{Jet quenching parameter in QCD kinetic theory}",
    eprint = "2312.00447",
    archivePrefix = "arXiv",
    primaryClass = "hep-ph",
    doi = "10.1103/PhysRevD.110.034019",
    journal = "Phys. Rev. D",
    volume = "110",
    number = "3",
    pages = "034019",
    year = "2024"
}

@article{Arnold:2002zm,
    author = "Arnold, Peter Brockway and Moore, Guy D. and Yaffe, Laurence G.",
    title = "{Effective kinetic theory for high temperature gauge theories}",
    eprint = "hep-ph/0209353",
    archivePrefix = "arXiv",
    doi = "10.1088/1126-6708/2003/01/030",
    journal = "JHEP",
    volume = "01",
    pages = "030",
    year = "2003"
}

@article{Kurkela:2014tea,
    author = "Kurkela, Aleksi and Lu, Egang",
    title = "{Approach to Equilibrium in Weakly Coupled Non-Abelian Plasmas}",
    eprint = "1405.6318",
    archivePrefix = "arXiv",
    primaryClass = "hep-ph",
    reportNumber = "CERN-PH-TH-2014-093",
    doi = "10.1103/PhysRevLett.113.182301",
    journal = "Phys. Rev. Lett.",
    volume = "113",
    number = "18",
    pages = "182301",
    year = "2014"
}

@article{Kurkela:2015qoa,
    author = "Kurkela, Aleksi and Zhu, Yan",
    title = "{Isotropization and hydrodynamization in weakly coupled heavy-ion collisions}",
    eprint = "1506.06647",
    archivePrefix = "arXiv",
    primaryClass = "hep-ph",
    reportNumber = "CERN-PH-TH-2015-142",
    doi = "10.1103/PhysRevLett.115.182301",
    journal = "Phys. Rev. Lett.",
    volume = "115",
    number = "18",
    pages = "182301",
    year = "2015"
}

@article{Fu:2021jhl,
    author = "Fu, Yu and Ghiglieri, Jacopo and Iqbal, Shahin and Kurkela, Aleksi",
    title = "{Thermalization of non-Abelian gauge theories at next-to-leading order}",
    eprint = "2110.01540",
    archivePrefix = "arXiv",
    primaryClass = "hep-ph",
    doi = "10.1103/PhysRevD.105.054031",
    journal = "Phys. Rev. D",
    volume = "105",
    number = "5",
    pages = "054031",
    year = "2022"
}

@article{Kurkela:2018xxd,
    author = "Kurkela, Aleksi and Mazeliauskas, Aleksas",
    title = "{Chemical Equilibration in Hadronic Collisions}",
    eprint = "1811.03040",
    archivePrefix = "arXiv",
    primaryClass = "hep-ph",
    reportNumber = "CERN-TH-2018-238",
    doi = "10.1103/PhysRevLett.122.142301",
    journal = "Phys. Rev. Lett.",
    volume = "122",
    pages = "142301",
    year = "2019"
}

@article{Kurkela:2018wud,
    author = {Kurkela, Aleksi and Mazeliauskas, Aleksas and Paquet, Jean-Fran{\c{c}}ois and Schlichting, S{\"o}ren and Teaney, Derek},
    title = "{Matching the Nonequilibrium Initial Stage of Heavy Ion Collisions to Hydrodynamics with QCD Kinetic Theory}",
    eprint = "1805.01604",
    archivePrefix = "arXiv",
    primaryClass = "hep-ph",
    doi = "10.1103/PhysRevLett.122.122302",
    journal = "Phys. Rev. Lett.",
    volume = "122",
    number = "12",
    pages = "122302",
    year = "2019"
}

@article{Du:2020zqg,
    author = {Du, Xiaojian and Schlichting, S{\"o}ren},
    title = "{Equilibration of the Quark-Gluon Plasma at Finite Net-Baryon Density in QCD Kinetic Theory}",
    eprint = "2012.09068",
    archivePrefix = "arXiv",
    primaryClass = "hep-ph",
    doi = "10.1103/PhysRevLett.127.122301",
    journal = "Phys. Rev. Lett.",
    volume = "127",
    number = "12",
    pages = "122301",
    year = "2021"
}

@article{Boguslavski:2024kbd,
    author = "Boguslavski, Kirill and Lindenbauer, Florian",
    title = "{Soft-gluon exchange matters: Isotropic screening in QCD kinetic theory}",
    eprint = "2407.09605",
    archivePrefix = "arXiv",
    primaryClass = "hep-ph",
    doi = "10.1103/PhysRevD.110.074017",
    journal = "Phys. Rev. D",
    volume = "110",
    number = "7",
    pages = "074017",
    year = "2024"
}

@article{BarreraCabodevila:2025ogv,
    author = "Barrera Cabodevila, Sergio and Kurkela, Aleksi and Lindenbauer, Florian",
    title = "{Solving the QCD effective kinetic theory with neural networks}",
    eprint = "2506.19632",
    archivePrefix = "arXiv",
    primaryClass = "hep-ph",
    doi = "10.1140/epjc/s10052-025-15040-w",
    journal = "Eur. Phys. J. C",
    volume = "85",
    number = "11",
    pages = "1304",
    year = "2025"
}

@article{Baier:2000sb,
    author = "Baier, R. and Mueller, Alfred H. and Schiff, D. and Son, D. T.",
    title = "{'Bottom up' thermalization in heavy ion collisions}",
    eprint = "hep-ph/0009237",
    archivePrefix = "arXiv",
    doi = "10.1016/S0370-2693(01)00191-5",
    journal = "Phys. Lett. B",
    volume = "502",
    pages = "51--58",
    year = "2001"
}

@software{kurkela_2023_10409474,
  author       = {Kurkela, Aleksi and
                  Lindenbauer, Florian},
  title        = {EKTqhat - Effective kinetic theory solver with jet
                   quenching parameter
                  },
  month        = dec,
  year         = 2023,
  publisher    = {Zenodo},
  doi          = {10.5281/zenodo.10409474},
  url          = {https://doi.org/10.5281/zenodo.10409474},
}

@article{Lappi:2011ju,
    author = "Lappi, T.",
    title = "{Gluon spectrum in the glasma from JIMWLK evolution}",
    eprint = "1105.5511",
    archivePrefix = "arXiv",
    primaryClass = "hep-ph",
    doi = "10.1016/j.physletb.2011.08.011",
    journal = "Phys. Lett. B",
    volume = "703",
    pages = "325--330",
    year = "2011"
}

@article{Mueller:1999pi,
    author = "Mueller, Alfred H.",
    title = "{The Boltzmann equation for gluons at early times after a heavy ion collision}",
    eprint = "hep-ph/9909388",
    archivePrefix = "arXiv",
    reportNumber = "CU-TP-950",
    doi = "10.1016/S0370-2693(00)00084-8",
    journal = "Phys. Lett. B",
    volume = "475",
    pages = "220--224",
    year = "2000"
}

@article{Ipp:2020nfu,
    author = {Ipp, Andreas and M{\"u}ller, David I. and Schuh, Daniel},
    title = "{Jet momentum broadening in the pre-equilibrium Glasma}",
    eprint = "2009.14206",
    archivePrefix = "arXiv",
    primaryClass = "hep-ph",
    doi = "10.1016/j.physletb.2020.135810",
    journal = "Phys. Lett. B",
    volume = "810",
    pages = "135810",
    year = "2020"
}

@article{Avramescu:2023qvv,
    author = {Avramescu, Dana and B{\u{a}}ran, Virgil and Greco, Vincenzo and Ipp, Andreas and M{\"u}ller, David. I. and Ruggieri, Marco},
    title = "{Simulating jets and heavy quarks in the glasma using the colored particle-in-cell method}",
    eprint = "2303.05599",
    archivePrefix = "arXiv",
    primaryClass = "hep-ph",
    doi = "10.1103/PhysRevD.107.114021",
    journal = "Phys. Rev. D",
    volume = "107",
    number = "11",
    pages = "114021",
    year = "2023"
}

@article{Barata:2024xwy,
    author = "Barata, Jo{\~a}o and Hauksson, Sigtryggur and Mayo L{\'o}pez, Xo{\'a}n and Sadofyev, Andrey V.",
    title = "{Jet quenching in the glasma phase: Medium-induced radiation}",
    eprint = "2406.07615",
    archivePrefix = "arXiv",
    primaryClass = "hep-ph",
    doi = "10.1103/PhysRevD.110.094055",
    journal = "Phys. Rev. D",
    volume = "110",
    number = "9",
    pages = "094055",
    year = "2024"
}

@article{Altenburger:2025iqa,
    author = "Altenburger, Alois and Boguslavski, Kirill and Lindenbauer, Florian",
    title = "{Jet broadening and radiation in the early anisotropic plasma in heavy-ion collisions}",
    eprint = "2509.03868",
    archivePrefix = "arXiv",
    primaryClass = "hep-ph",
    reportNumber = "MIT-CTP/5917",
    month = "9",
    year = "2025"
}

@article{Boguslavski:2024ezg,
    author = "Boguslavski, Kirill and Kurkela, Aleksi and Lappi, Tuomas and Lindenbauer, Florian and Peuron, Jarkko",
    title = "{Jet momentum broadening during initial stages in heavy-ion collisions}",
    eprint = "2303.12595",
    archivePrefix = "arXiv",
    primaryClass = "hep-ph",
    doi = "10.1016/j.physletb.2024.138525",
    journal = "Phys. Lett. B",
    volume = "850",
    pages = "138525",
    year = "2024"
}

@article{Zhou:2024ysb,
    author = "Zhou, Fabian and Brewer, Jasmine and Mazeliauskas, Aleksas",
    title = "{Minijet quenching in non-equilibrium quark-gluon plasma}",
    eprint = "2402.09298",
    archivePrefix = "arXiv",
    primaryClass = "hep-ph",
    doi = "10.1007/JHEP06(2024)214",
    journal = "JHEP",
    volume = "06",
    pages = "214",
    year = "2024"
}

@article{Barata:2021wuf,
    author = "Barata, Jo{\~a}o and Mehtar-Tani, Yacine and Soto-Ontoso, Alba and Tywoniuk, Konrad",
    title = "{Medium-induced radiative kernel with the Improved Opacity Expansion}",
    eprint = "2106.07402",
    archivePrefix = "arXiv",
    primaryClass = "hep-ph",
    doi = "10.1007/JHEP09(2021)153",
    journal = "JHEP",
    volume = "09",
    pages = "153",
    year = "2021"
}

@article{Mehtar-Tani:2019ygg,
    author = "Mehtar-Tani, Yacine and Tywoniuk, Konrad",
    title = "{Improved opacity expansion for medium-induced parton splitting}",
    eprint = "1910.02032",
    archivePrefix = "arXiv",
    primaryClass = "hep-ph",
    doi = "10.1007/JHEP06(2020)187",
    journal = "JHEP",
    volume = "06",
    pages = "187",
    year = "2020"
}

@article{Arnold:2008iy,
    author = "Arnold, Peter Brockway",
    title = "{Simple Formula for High-Energy Gluon Bremsstrahlung in a Finite, Expanding Medium}",
    eprint = "0808.2767",
    archivePrefix = "arXiv",
    primaryClass = "hep-ph",
    doi = "10.1103/PhysRevD.79.065025",
    journal = "Phys. Rev. D",
    volume = "79",
    pages = "065025",
    year = "2009"
}

@article{Boguslavski:2026wma,
    author = {Boguslavski, Kirill and H{\"o}rl, Lucas and Lindenbauer, Florian},
    title = "{Jet energy loss in anisotropic plasmas meets limiting attractors}",
    eprint = "2603.09912",
    archivePrefix = "arXiv",
    primaryClass = "hep-ph",
    reportNumber = "MIT-CTP/6013",
    doi = "10.1103/69px-pnlm",
    journal = "Phys. Rev. D",
    volume = "114",
    number = "5",
    pages = "054019",
    year = "2026"
}

@article{Lindenbauer:2025ctw,
    author = "Lindenbauer, Florian",
    title = "{Gluon splitting rates in an anisotropic plasma in the AMY formalism}",
    eprint = "2509.09897",
    archivePrefix = "arXiv",
    primaryClass = "hep-ph",
    reportNumber = "MIT-CTP/5920",
    month = "9",
    year = "2025"
}

@article{Barata:2024bqp,
    author = "Barata, Jo{\~a}o and Salgado, Carlos A. and Silva, Jo{\~a}o M.",
    title = "{Gluon to $ \textrm{q}\overline{\textrm{q}} $ antenna in anisotropic QCD matter: spin-polarized and azimuthal jet observables}",
    eprint = "2407.04774",
    archivePrefix = "arXiv",
    primaryClass = "hep-ph",
    doi = "10.1007/JHEP12(2024)023",
    journal = "JHEP",
    volume = "12",
    pages = "023",
    year = "2024"
}

@misc{lindenbauer_2025_17552870,
  author       = {Lindenbauer, Florian},
  title        = {Improved Opacity Expansion Solver for Time-
                   Dependent Matter
                  },
  month        = nov,
  year         = 2025,
  publisher    = {Zenodo},
  doi          = {10.5281/zenodo.17552870},
  url          = {https://doi.org/10.5281/zenodo.17552870},
}

@article{Apolinario:2022vzg,
    author = "Apolin{\'a}rio, Liliana and Lee, Yen-Jie and Winn, Michael",
    title = "{Heavy quarks and jets as probes of the QGP}",
    eprint = "2203.16352",
    archivePrefix = "arXiv",
    primaryClass = "hep-ph",
    doi = "10.1016/j.ppnp.2022.103990",
    journal = "Prog. Part. Nucl. Phys.",
    volume = "127",
    pages = "103990",
    year = "2022"
}

\end{document}